# Selective reflection spectroscopy of a vapour at a calcium fluoride interface

T. Passerat de Silans[(1)], A. Laliotis[(1)], M. Romanelli[(1)], P. Chaves de Souza Segundo[(1)], I. Maurin[(1)], D. Bloch[(1)], M. Ducloy[(1)],
[(1)]Laboratoire de Physique des Lasers, UMR 7538 du CNRS et de l'Université Paris 13, 99 avenue Jean-Baptiste Clément, 93430 Villetaneuse, France
A. Sarkisyan[(2)], D. Sarkisyan[(2)]
[(2)] Institute for Physical Research, Armenian Academy of Sciences, Ashtarak 2, Armenia

**Abstract.** Fluoride materials exhibit surface resonances located in the thermal infrared. This makes them interesting to search for a fundamental temperature dependence of the atom-surface interaction, originating in the near-field thermal emissivity of the surface. Preliminary selective reflection experiments performed on a special Cs vapour cell that includes a $CaF_2$ interface show a temperature dependence, yet to be analyzed.

## 1. INTRODUCTION

It was shown that the universal van der Waals (vW) type long-range attraction between an atom and a surface is susceptible to be resonantly enhanced when an atomic virtual emission couples to a virtual absorption into a surface polariton mode. This resonant coupling can even turn the attraction into a repulsion [1]. Such a process is restricted to excited atoms, as they must be emitters. The process can however be extended to the inverse coupling between an atomic absorption and a surface emission, provided that the surface is thermally excited as the considered resonant process couples a single mode of atomic excitation, and a continuum of modes for the surface excitation. The detailed theory of this reverse coupling has just appeared [2]. The temperature dependence will be strong when surface resonances are located in the thermal infrared, and when the considered atomic states exhibit a strong (dipole) coupling in this infrared domain. This justifies our interest for selective reflection spectroscopy (SR) of a vapour at fluoride interface : (i) fluoride windows, such as $CaF_2$, $BaF_2$, $MgF_2$, transparent in the visible domain, exhibit surface resonances in the far infrared domain [3] respectively at ~24.0 µm, 32 µm and 18.5 µm; (ii) SR spectroscopy [1] probes the optical response of a vapour at a distance ~$\lambda/2\pi$ (≤ 100 nm) from the interface, well suited for the study of the atom-surface interaction. The considered experiment aims at the excited Cs(8P) level, because of its strong virtual absorption to Cs(7D) (at 29, 36, 39 µm, depending on the considered fine structure levels). The non-contact interaction between Cs(8P) and the hot $CaF_2$ window is hence predicted to depend strongly on temperature, and should be amenable to a repulsion.

## 2. TESTING A CELL SPECIALLY DESIGNED FOR UNUSUAL INTERFACES

The Armenian team has fabricated a special vapour cell resistant to hot Cs vapour that includes a fluoride window. This was a challenge because fluoride materials are mechanically fragile, sensible to thermal shocks, and cannot be glued, even with a mineral glue, to a sapphire tubing because of too different thermal expansions. This is why the cell is made of a standard all-sapphire cell (fig. 1), with a very thick fluoride tube inside, in near contact with one of the sapphire window. This allows a large thermal gradient inside the fluoride window, one side of the sapphire tubing remaining at ambient temperature (left side in fig.1). With the strong dependence of the Cs density with temperature, the absorption in the interstitial region (<100 µm) between the sapphire (thin) window and the (thick) fluoride window is in principle negligible, so that optically the reflection at the interface between the fluoride window (presently $CaF_2$) and the Cs vapour directly yields the SR signal of interest. Note that the cell design allows one to conveniently perform simultaneous SR experiments on both fluoride and sapphire interfaces, under identical experimental conditions for the Cs vapour.

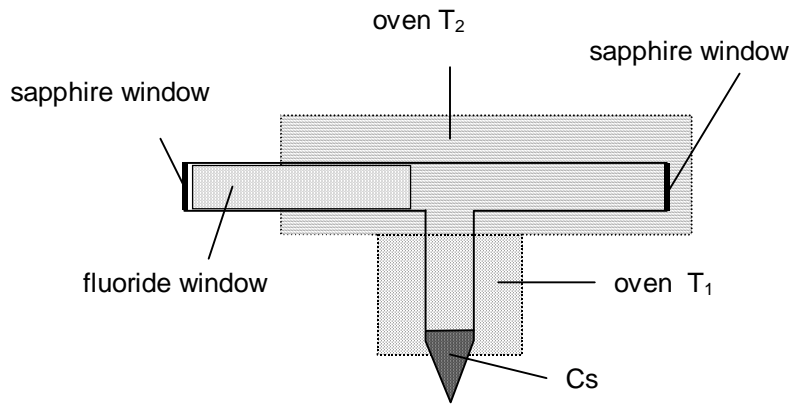

Figure 1 : *Scheme of the vapour cell with fluoride window. One facet of the 8 cm-thick fluoride window remains outside the oven, at ambient temperature, while its internal facet is at $T_2 > T_1$, $T_1$ being the temperature governing the Cs density.*

The cell was initially tested on the well-known $D_1$ resonance line of Cs ($\lambda$=894 nm), when no special far IR coupling occurs. This was particularly needed because of impurities in the prototype cell that we have tested : owing to a large amount of velocity-changing-collisions, one has indeed observed a considerable broadening -and even shift- for a standard saturated absorption spectrum. The validity of the cell design could be however explored because one notable advantage of *linear* SR spectroscopy is that it is nearly insensitive to velocity-changing-collisions : for low Cs density, only a minor broadening (~20MHz) -relatively to the natural width (5 MHz)- was observed, along with a ~ -5MHz shift. In spite of this, and of the fact that the vW interaction exerted onto the $D_1$ line is always weak, leading only to minor distortion and shift of the lineshape, we have been able to extract [5] the respective vW interaction exerted by the two different windows. Technically, the precision of the evaluation [1] was improved by a simultaneous fitting of the two spectra, the constraint being that the uncontrolled broadening and shift induced by the impurities (*i.e.* additional adjustable parameters in the fitting process) must be the same for both windows. On this prototype cell, we have established that the vW interaction is weaker for $CaF_2$ than for sapphire by a factor ~0.6 ± 0.25, as expected from the respective refractive index values of these materials (sapphire : ~1.76, $CaF_2$ : ~1.43).

## 3. TOWARDS A TEMPERATURE-DEPENDENT SURFACE INTERACTION

We are now starting to investigate the SR spectrum on the $6S_{1/2} \rightarrow 8P_{3/2}$ transition (λ= 388 nm) on a cell with a $CaF_2$ interface, in order to explore the predicted temperature dependence of the vW interaction. SR spectroscopy on this line had already been investigated at the interface with a sapphire window, for which no surface temperature dependence is expected [6]. With the prototype cell and its impurities, we could simultaneously record the SR spectra on this transition for both windows. However, the broadening originating in the impurities lowers the signal down so much, that the maximal temperature must be used, jeopardizing a study of the temperature dependence. Very recently, a second cell with a $CaF_2$ window was implemented and tested, revealing a negligible amount of impurities. Preliminary results, yet to be analyzed, show large changes of the SR spectra with temperature at the interface with $CaF_2$, and no changes at a sapphire interface (see fig. 2). Refined theoretical predictions are now expected with a study, performed by P. Echegut at CRMHT-Orléans, of the optical properties in the thermal IR range of fluoride materials between 20°C and 500°C.

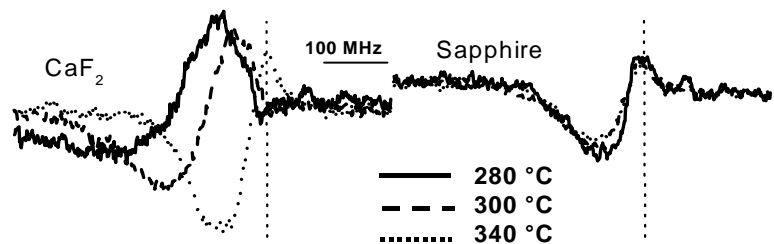

Figure 2 : *Simultaneously recorded SR spectra on the 388 nm line at a $CaF_2$ and sapphire interfaces at various window temperatures $T_2$ . The Cs reservoir is maintained at 200 °C. The vertical dashed line is a frequency marker, corresponding to the free-space atomic transition.*

**References**

[1] H. Failache *et al.,* Eur. Phys. J. D **23**, 237 (2003) et Phys. Rev. Lett. **83**, 5467 (1999)
[2] M.-P. Gorza and M. Ducloy, Eur. Phys. J. D **40**, 343 (2006)
[3] S. Saltiel, D. Bloch and M. Ducloy, Opt. Commun., **265**, 220-233 (2006)
[4] G. Dutier *et al.* in "Laser Spectroscopy - Proceedings of the XVI International Conference" (P. Hannaford *et al.* eds.), pp 277-284, World Scientific, Singapore (2004)
[5] A. Laliotis *et al.* , in preparation
[6] P. Chaves de Souza Segundo *et al.*, Laser Phys. **17**, 983 (2007)